\documentclass{PoS}
\usepackage{amsmath}
\allowdisplaybreaks

\title{Rare {\em B} decays}

\ShortTitle{Rare $B$ decays}

\author{\speaker{David M. Straub}\\
        Scuola Normale Superiore and INFN, Piazza dei Cavalieri 7, 56126 Pisa, Italy\\
        E-mail: \email{david.straub@sns.it}}

\abstract{Rare radiative, leptonic and semileptonic $B$ meson decays are valuable probes of physics beyond the Standard Model. This talk reviews the sensitivity to new physics of observables related to two rare $B$ decays to be measured in the near future at LHC, $B_s\to\mu^+\mu^-$ and $B\to K^*\mu^+\mu^-$.}

\FullConference{The 13th International Conference on B-Physics at Hadron Machines - Beauty2011,\\
		April 04-08, 2011\\
		Amsterdam, The Netherlands}

\begin{document}

\section{Introduction}

\noindent
Rare $B$ decays are powerful probes of physics beyond the Standard Model (SM). Being induced by flavour-changing neutral currents, their branching ratios are strongly suppressed in the SM and sensitive to new physics (NP). CP asymmetries can probe non-standard CP violation and angular distributions of multi-body decays can be used to probe the chirality structure of the fundamental interactions. Beyond the SM, rare $B$ decays can thus contribute to understanding the NP flavour structure and eventually the origin of flavour.

A selection of observables in decays with a $b\to s$ FCNC transition is shown in table~\ref{tab:exp}. The first class of decays, inclusive radiative and semileptonic decays, already provides strong constraints on models beyond the SM. Their branching ratios can be predicted with comparatively small theoretical uncertainties. Experimentally, they require the clean environment of $e^+$-$e^-$ colliders to be measured, so significant improvement is not expected before the advent of the Super $B$ factories. The same is true for the second class of decays, the ones with a neutrino pair in the final state, which have not been observed yet but would be valuable probes of NP \cite{Altmannshofer:2009ma}.
In the early LHC era, the most promising rare $B$ decays to look for NP are the third class in table~\ref{tab:exp}, {\em exclusive} decays with a muon pair in the final state.

\begin{table}[tbp]
 \centering
 \renewcommand{\arraystretch}{1.2}
\begin{tabular}{llll} \hline
 mode & SM & exp. 
\\\hline
BR$(B\to X_s\gamma)$  & $3.2\times 10^{-4}$ & $(3.52\pm0.25)\times 10^{-4}$ \\
$A_\text{CP}(B\to X_s\gamma)$& $\sim1\%$ & $(-1.2\pm2.8)\%$ \\
BR$(B\to X_s\ell^+\ell^-)_{\text{low} q^2}$ & $1.6\times 10^{-6}$ & $(1.59\pm0.49)\times 10^{-6}$ \\
\hline
BR$(B\to K\nu\bar\nu)$ & $4\times 10^{-6}$ & $< 14\times 10^{-6}$ \\
BR$(B\to K^*\nu\bar\nu)$ & $6.8\times 10^{-6}$ & $< 80\times 10^{-6}$ \\
 $F_L(B\to K^*\nu\bar\nu)$ & $0.54$ &  -- \\
\hline
BR$(B_s\to \mu^+\mu^-)$ & $3.2\times 10^{-9}$ & $< 43\times 10^{-9}$ \\    
BR$(B_d\to \mu^+\mu^-)$ & $1.0\times 10^{-10}$ & $< 76\times 10^{-10}$ \\
$A_\text{FB}(B\to K^* \mu^+\mu^-)_{\text{low} q^2}$ & $0.03$ & $0.42\pm0.37$ \\
\hline
\end{tabular}
\caption{Approximate SM predictions and current experimental bounds for a selection of rare $B$ decay observables (see \cite{Asner:2010qj} and references therein). The first two classes, inclusive decays and decays with a neutrino pair in the final state, are difficult to measure at hadron colliders.}
\label{tab:exp}
\end{table}

Contributions from physics beyond the SM can be encoded in the Wilson coefficients of local operators by means of the operator product expansion. For decays with a $b\to s\ell^+\ell^-$ transition, the part of the effective Hamiltonian most sensitive to NP effects reads
\begin{equation} \label{eq:HeffNP}
    {\mathcal H}_{\text{eff}} = - \frac{4\,G_F}{\sqrt{2}}V_{tb}V_{ts}^*
\sum_{i=7,8,9,10,P,S} (C_i \mathcal O_i + C'_i \mathcal
O'_i)\,,
\end{equation}
where the relevant operators are given by
\begin{align}
{\mathcal{O}}_{7} &= \frac{e^2}{16\pi^2} m_b
(\bar{s} \sigma_{\mu \nu} P_R b) F^{\mu \nu} ,&
{\mathcal{O}}_{8} &= \frac{g_3^2}{16\pi^2} m_b
(\bar{s} \sigma_{\mu \nu} T^a P_R b) G^{\mu \nu \, a} ,&
\label{eq:DF1-1}
\\
{\mathcal{O}}_{9} &= \frac{e^2}{16\pi^2} 
(\bar{s} \gamma_{\mu} P_L b)(\bar{\ell} \gamma^\mu \ell) ,&
{\mathcal{O}}_{10} &=\frac{e^2}{16\pi^2} 
(\bar{s}  \gamma_{\mu} P_L b)(  \bar{\ell} \gamma^\mu \gamma_5 \ell) ,&
\label{eq:DF1-2}
\\
{\mathcal{O}}_{S} &=\frac{e^2}{16\pi^2}
m_b (\bar{s} P_R b)(  \bar{\ell} \ell) ,&
{\mathcal{O}}_{P} &=\frac{e^2}{16\pi^2}
m_b (\bar{s} P_R b)(  \bar{\ell} \gamma_5 \ell) ,&
\label{eq:DF1-3}
\end{align}
and the chirality-flipped operators $\mathcal O_i'$ are obtained from (\ref{eq:DF1-1})--(\ref{eq:DF1-3}) by the replacement $P_L\leftrightarrow P_R$. In the SM, the scalar and pseudoscalar operators $\mathcal O_{S,P}$ as well as all the ``primed'' operators $\mathcal O_i'$ have negligible Wilson coefficients.

While the short-distance contributions are straightforward to calculate in the SM and a given NP model, the long-distance hadronic physics governed by the SM constitute a considerable source of uncertainty in exclusive decays and will be discussed in turn for the two decays of interest.

\section{$B_{s,d}\to\mu^+\mu^-$}

\noindent
The decays $B_{q}\to\mu^+\mu^-$, where $q=s,d$, are strongly helicity suppressed in the SM.
Due to the purely leptonic final state, the hadronic uncertainties are limited to the $B_q$ meson decay constant $f_{B_q}$. Within the SM, the uncertainties can be further reduced by making use of the experimental measurement of $\Delta M_q$ to trade the decay constant squared for the bag parameter $\hat B_q$, which is known more precisely \cite{Buras:2003td}. One then arrives at
\begin{align}\label{eq:Bsmumu-SM}
 \text{BR}(B_s\to\mu^+\mu^-)_\text{SM} &= (3.2 \pm 0.2) \times 10^{-9},\\
 \text{BR}(B_d\to\mu^+\mu^-)_\text{SM} &= (0.10 \pm 0.01) \times 10^{-9}.
\end{align}
Experimentally, neither mode has been observed yet and the current 95\% C.L. upper bounds still lie one/two orders of magnitude above the SM \cite{CDFPubNote9892}:
\begin{align}\label{eq:Bsmumu-exp}
 \text{BR}(B_s\to\mu^+\mu^-)_\text{exp} &< 43 \times 10^{-9},\\
 \text{BR}(B_d\to\mu^+\mu^-)_\text{exp} &< 7.6 \times 10^{-9}.
\end{align}
However, as the Tevatron experiments are improving their bounds, LHCb is closing in quickly \cite{Aaij:2011rj} and will set a stronger bound -- or observe NP --  by the end of 2011.

\begin{figure}[tbp]
\centering
\parbox[b]{0.53\textwidth}{
\includegraphics[width=0.48\textwidth]{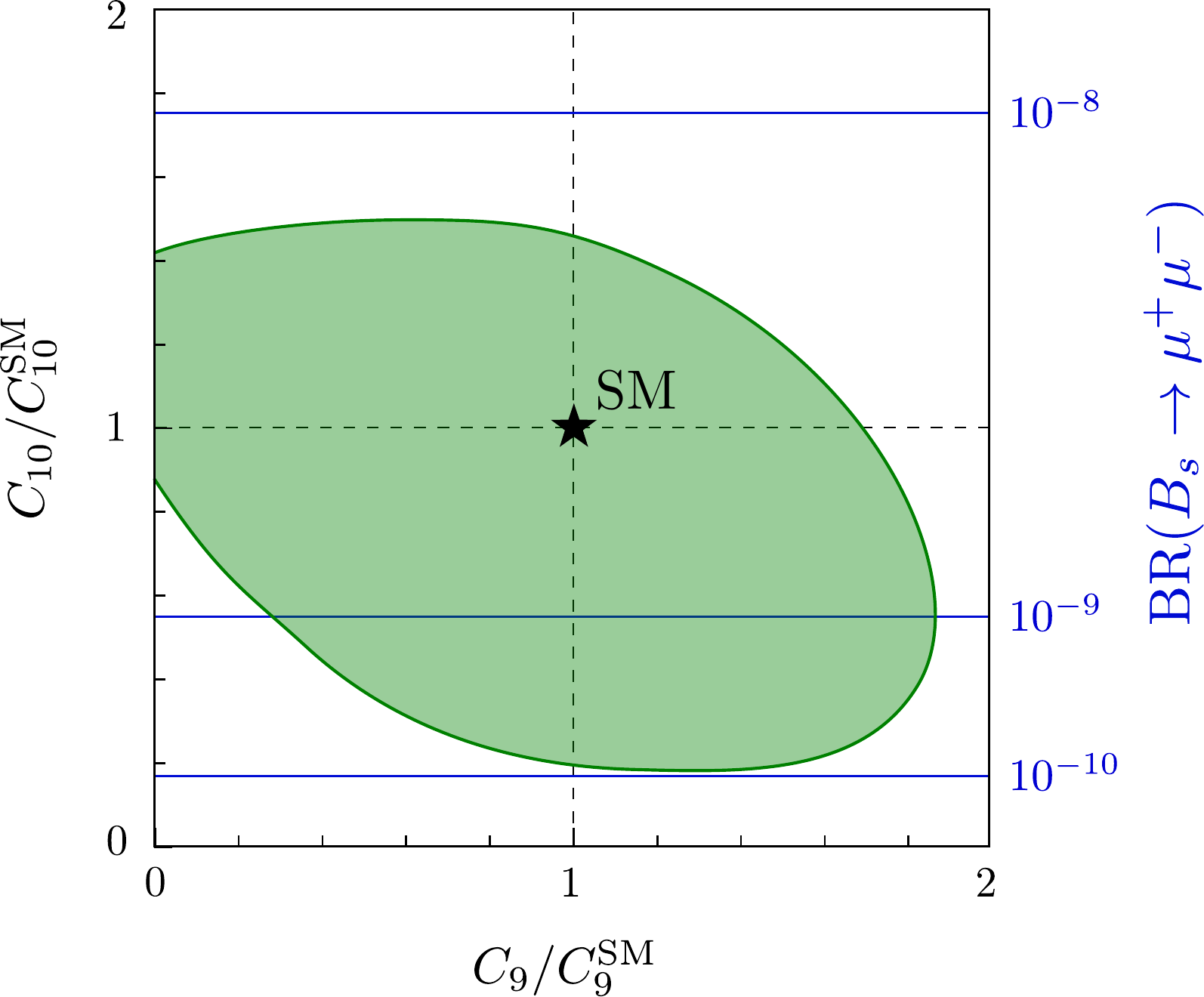}\hfill
}
\parbox[b]{0.42\textwidth}{
\caption{Sketch of the allowed area at 95\% C.L. in the $C_9$-$C_{10}$ plane taking into account data on $b\to s\gamma$ and inclusive as well as exclusive $b\to s\ell^+\ell^-$ decays (based on ref. \cite{Bobeth:2010wg}). The right-hand axis shows the corresponding value of $\text{BR}(B_s\to\mu^+\mu^-)$ (neglecting its theoretical uncertainty).}
\label{fig:c9c10}
}
\end{figure}

In a generic NP model, the branching ratio is given by
\begin{equation}
 \text{BR}(B_q\to\mu^+\mu^-) = \tau_{B_q} f_{B_q}^2 m_{B_q}
\frac{\alpha_\text{em}^2 G_F^2 }{16 \pi^3} 
 |V_{tb} V_{tq}^*|^2 \sqrt{ 1 - \frac{4 m_\mu^2}{m_{B_q}^2}} \left[|S|^2 \left( 1 - \frac{4 m_\mu^2}{m_{B_q}^2} \right) + |P|^2\right],
\label{eq:BRBsmumu1}
\end{equation}
where\footnote{In (\ref{eq:BRBsmumu2}), $C_i^{(s)}\equiv C_i$ are the Wilson coefficients of the operators in (\ref{eq:DF1-1})--(\ref{eq:DF1-3}), while $C_i^{(d)}$ are the ones of the corresponding $b\to d$ operators.}
\begin{equation}
S = \frac{m_{B_q}^2}{2} (C_S^{(q)} - C_S^{\prime(q)}) , \qquad
P = \frac{m_{B_q}^2}{2} (C_P^{(q)} - C_P^{\prime(q)}) + m_\mu (C_{10}^{(q)} - C_{10}^{\prime(q)}).
\label{eq:BRBsmumu2}
\end{equation}
In models where NP enters $B_s\to\mu^+\mu^-$ only through the SM Wilson coefficient $C_{10}$, an order-of-magnitude enhancement of the branching ratio is disfavoured due to constraints from inclusive and exclusive $b\to s\ell^+\ell^-$ transitions on $C_{10}$ (see the sketch in figure~\ref{fig:c9c10}). In that case, BR$(B_s\to\mu^+\mu^-)\lesssim10^{-8}$ \cite{Bobeth:2010wg}.

Much larger enhancements are possible in principle in models with contributions to the scalar and/or pseudoscalar Wilson coefficients $C_{S,P}$. In two-Higgs-doublet models, a neutral Higgs penguin contributes to the branching fraction with an enhancement factor of $\tan\beta^4$, where $\tan\beta$ is the ratio of Higgs VEVs. In the MSSM, this dependence is even $\tan\beta^6$. Consequently, an upper bound  $\text{BR}(B_s\to\mu^+\mu^-) <  10^{-8}$ would already constrain numerous well-motivated NP scenarios, such as SUSY GUTs with a unification of Yukawa couplings \cite{Altmannshofer:2008vr}.\footnote{On the other hand, even within the MFV MSSM very large $\tan\beta$ would remain a valid possibility if the trilinear couplings are small, such as in gauge mediation scenarios \cite{Altmannshofer:2010zt}.}

An important test of the principle of Minimal Flavour Violation (MFV, \cite{Buras:2000dm}) is represented by the measurement of the ratio of the two $B_q\to\mu^+\mu^-$ branching ratios. In MFV, one has $C_i^{(s)}=C_i^{(d)}$ in (\ref{eq:BRBsmumu2}) and consequently model-independently
\begin{equation}
\frac{\text{BR}(B_s\to\mu^+\mu^-)}{\text{BR}(B_d\to\mu^+\mu^-)} =
\frac{\tau_{B_s}f_{B_s}^2m_{B_s}}{\tau_{B_d}f_{B_d}^2m_{B_d}}
\left|\frac{V_{ts}}{V_{td}}\right|^2.
\label{eq:sd}
\end{equation}
While a simultaneous enhancement over the SM satisfying (\ref{eq:sd}) would be a strong indication in favour of MFV, a non-SM effect incompatible with (\ref{eq:sd}), which is predicted in various NP models (see \cite{Straub:2010ih} for a comparison), would immediately rule out MFV.

\section{$B\to K^*\mu^+\mu^-$}

\noindent
The exclusive decay $B\to K^*\mu^+\mu^-$ is more involved theoretically as well as experimentally compared to the previous decays, but it also allows to probe more diverse NP effects. It is sensitive to all the operators in (\ref{eq:DF1-1})--(\ref{eq:DF1-3}); the angular distribution of the all-charged four-body final state $\bar B\to \bar K^{*0}(\to K^-\pi^+)\mu^+\mu^-$ gives access to many observables potentially sensitive to NP; and the charge conjugated mode $B\to K^{*0}(\to K^+\pi^-)\mu^+\mu^-$, which can be distinguished from the former just by means of the meson charges, allows a straightforward measurement of CP asymmetries.

On the theory side, the decay poses several challenges. First, it requires the calculation by non-perturbative methods of the 7 $B\to K^*$ form factors, which are functions of the dilepton invariant mass-squared $q^2$. Second, at intermediate $q^2$, resonant charmonium production $B\to K^*\psi(\to\ell^+\ell^-)$ leads to a breakdown of quark-hadron duality. Third, there are additional non-factorizable strong interaction effects that cannot be expressed in terms of form factors. (See e.g. \cite{Khodjamirian:2010vf} for a recent discussion.)

At low $q^2$, i.e. below the charmonium resonances, QCD factorization can be used in the heavy quark limit, which reduces the number of independent form factors from 7 to 2 and allows a systematic calculation of non-factorizable corrections \cite{Beneke:2001at}. The remaining theoretical uncertainties then reside in phenomenological parameters like meson distribution amplitudes, in the form factors themselves, as well as in possible corrections of higher order in the ratio $\Lambda_\text{QCD}/m_b$. Two different approaches of treating these power-suppressed corrections have been followed in the literature: In \cite{Altmannshofer:2008dz}, QCD sum rules on the light cone (LCSR) have been used to obtain all the 7 $B\to K^*$ form factors without the need to resort to the heavy quark limit, thereby taking into account one source of $\Lambda_\text{QCD}/m_b$ corrections. Results based on LCSR are then used to argue that the remaining $\Lambda_\text{QCD}/m_b$ corrections are expected to be of $O(\alpha_s)$. In \cite{Egede:2008uy}, none of the $\Lambda_\text{QCD}/m_b$ corrections have been included in the calculation, but a generous additional theory uncertainty has been assumed instead.

The high-$q^2$ region above the charmonium resonances has recently attracted increasing attention from the theory community \cite{Bobeth:2010wg,Beylich:2011aq}. While QCD factorization and LCSR methods are not applicable in this kinematical domain, a local operator product expansion in powers of $1/\sqrt{q^2}$ allows a systematic calculation of the observables \cite{Grinstein:2004vb,Beylich:2011aq}. In \cite{Beylich:2011aq}, it has been argued that, in contrast to the low-$q^2$ region, non-perturbative corrections {\em not} accounted for by the form factors are small. A drawback is the invalidity of LCSR at high $q^2$ necessitating an interpolation of the form factors from low to high $q^2$, but lattice calculations of the form factors are ongoing and might improve the situation \cite{Liu:2011ra}.

\begin{figure}[tbp]
\centering
\includegraphics[width=0.45\textwidth]{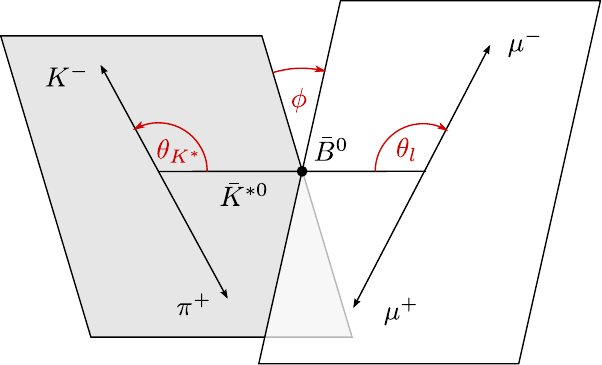}
\caption{Kinematical angles in the $\bar B\to \bar K^{*0}(\to K^-\pi^+)\mu^+\mu^-$ decay.}
\label{fig:bsll}
\end{figure}

The full set of observables accessible in the angular distribution of the decay and its CP-conjugate is given by the angular coefficient functions $I_i(q^2)$ and $\bar I_i(q^2)$. Neglecting scalar operator contributions (which are strongly constrained by $B_s\to\mu^+\mu^-$ discussed above) and lepton mass effects (which is a good approximations for electrons and muons), there are 9+9 independent angular coefficients. They can be expressed in terms of transversity amplitudes, which are functions of the form factors and Wilson coefficients.
While the overall normalization of the angular coefficients is subject to considerable uncertainties, theoretically cleaner observables are obtained by normalizing them to the total invariant mass distribution. Furthermore, it makes sense to separate the observables into CP asymmetries $A_i$ and CP-averaged ones $S_i$. One thus arrives at \cite{Altmannshofer:2008dz}
\begin{equation}
 S_i = \left( I_i + \bar I_i \right) \bigg/ \frac{d(\Gamma+\bar\Gamma)}{dq^2} \,,
\qquad
 A_i = \left( I_i - \bar I_i \right) \bigg/ \frac{d(\Gamma+\bar\Gamma)}{dq^2}\,.
\label{eq:As}
\end{equation}

Not all of the $S_i$ and $A_i$ are both theoretically interesting and experimentally promising. Let us therefore focus on six ``highlight'' observables in the low $q^2$ region which are particularly sensitive to NP and which have good prospects of being measured with some precision in the near future.

At $B$ factories, at the Tevatron and at the early LHC, statistics is quite limited so a full angular analysis is not possible. Then, one can consider one- or two-dimensional angular distributions depending on a limited set of observables.
For example, the one-dimensional distributions in the angles $\theta_l$ and $\theta_{K^*}$ shown in figure \ref{fig:bsll} read
\begin{align}
\frac{d(\Gamma+\bar\Gamma)}{d\cos\theta_l \,dq^2} \bigg/ \frac{d(\Gamma+\bar\Gamma)}{dq^2}
&=
\frac{3}{8} \left[
(1+S_2) \left(1+\cos^2\theta_l\right)-2S_2\left(1-\cos^2\theta_l\right)
+2A_6 \cos\theta_l\right]
\,,\label{eq:1Dl}\\
\frac{d(\Gamma+\bar\Gamma)}{d\cos\theta_{K^*} \,dq^2} \bigg/ \frac{d(\Gamma+\bar\Gamma)}{dq^2}
&=
\frac{3}{4} \left[
(1+S_2) \left(1-\cos^2\theta_{K^*}\right)-2S_2\cos^2\theta_{K^*}
\right]
\,,\label{eq:1DK}
\end{align}
while $S_6$ (and $A_2$) can be obtained from the distribution (\ref{eq:1Dl}) with $(\Gamma+\bar\Gamma)\to(\Gamma-\bar\Gamma)$ in the numerator on the left-hand side\footnote{The appearance of the forward-backward CP asymmetry $A_6$ in the CP averaged angular distribution is linked to the definition of $\theta_l$, which is defined as the decay angle of the $\mu^-$ in the $B$ and the $\bar B$ decay here. Note that in \cite{:2008ju,:2009zv,Aaltonen:2011cn}, $\theta_l$ is the angle of $\mu^+$ ($\mu^-$) in the $B$ ($\bar B$) decay.}.
The most interesting observables here to probe NP are $S_6$, which is the forward-backward asymmetry, and $S_2$, which is the $K^*$ longitudinal polarization fraction (see the comparison of notations in table~\ref{tab:obs}).
First results on these distributions have been published by BaBar \cite{:2008ju}, Belle \cite{:2009zv} and CDF \cite{Aaltonen:2011cn}.

Two additional interesting observables could be obtained from the one-dimensional distribution in the angle $\phi$,
\begin{equation}
\frac{d(\Gamma+\bar\Gamma)}{d\phi \,dq^2} \bigg/ \frac{d(\Gamma+\bar\Gamma)}{dq^2}
=
\frac{1}{2\pi} \left[
 1  +
 S_3  \cos(2\phi) +
 A_9  \sin(2\phi)
\right]
\,.
\label{eq:1Dphi}
\end{equation}
While both the CP averaged coefficient $S_3$ and the CP asymmetry $A_9$ are negligible small in the SM, they could be nonzero in NP models with right handed currents since they are sensitive to the Wilson coefficients of the ``primed'' operators.

Finally, two additional observables which might be accessible even during the early LHC running are the observable $S_5$ and the CP-asymmetry $A_7$. They can both be extracted from the two-dimensional angular distribution in $\theta_{K^*}$ and $\phi$,
\begin{align}
\frac{d(\Gamma-\bar\Gamma)}{d\cos\theta_{K^*}d\phi \,dq^2} \bigg/ \frac{d(\Gamma+\bar\Gamma)}{dq^2}
&=
\frac{3}{64 \pi }
\big[3 \pi  \sin (2 \theta_{K^*}) (S_5 \cos \phi+A_7 \sin \phi)
\label{eq:2D}\\\nonumber
&+8 \sin
   ^2\theta_{K^*} (S_9 \sin (2 \phi )+A_2+A_3 \cos (2 \phi )+1)-16
   A_2 \cos ^2\theta_{K^*}
\big].
\end{align}
The observables in the second line of (\ref{eq:2D}) can be obtained from the one-dimensional distributions (\ref{eq:1Dl})--(\ref{eq:1Dphi}) with $(\Gamma+\bar\Gamma)\to(\Gamma-\bar\Gamma)$ in the numerator on the left-hand side, but they are negligibly small in the SM and are unlikely to be enhanced in the presence of NP \cite{Altmannshofer:2008dz}.
In \cite{Bharucha:2010bb}, it has been shown that even with an integrated luminosity of only $2~\text{fb}^{-1}$, LHCb can measure $S_5$ with a precision that already allows to probe certain NP scenarios. Since $A_7$ is accessible from the same distribution (\ref{eq:2D}), the naive expectation is that the sensitivity should be comparable.
Just as $A_9$, $A_7$ is a T-odd CP asymmetry, meaning that it is not suppressed by small strong phases \cite{Bobeth:2008ij}.
Sizable effects in $A_7$ of up to 20\% are expected in well-motivated NP scenarios, like the MFV MSSM or the MSSM with hierarchical sfermions and flavour-blind phases \cite{Altmannshofer:2009ne}.

\begin{table}
\begin{center}
\renewcommand{\arraystretch}{1.2}
\begin{tabular}{c|cccc|l}
Obs.  & \cite{Altmannshofer:2008dz} & \cite{Egede:2008uy} & \cite{Bobeth:2008ij} & \cite{:2008ju,:2009zv,Aaltonen:2011cn} & most sensitive to\\
\hline
$S_2$ & $S_2^c$ & $-F_L$ && $-F_L$ & $C_{7,9,10}^{(\prime)}$ \\
$S_6$ & $S_6^s$ & $\frac{4}{3}A_\text{FB}$ & $\frac{4}{3}A_\text{FB}$ & $-\frac{4}{3}A_\text{FB}$ & $C_7, C_9$ \\
$S_3$ & $S_3$ & $\frac{1}{2}(1-F_L)A_T^{(2)}$ &&  & $C_{7,9,10}^{\prime}$\\
$A_9$ & $A_9$ && $\frac{2}{3}A_9$ &  & $C_{7,9,10}^{\prime}$ \\
$S_5$ & $S_5$ &&  &  & $C_7,C_7',C_9,C_{10}'$\\
$A_7$ & $A_7$ && $-\frac{2}{3}A_7^D$ &  & $C_{7,10}^{(\prime)}$
\end{tabular}
\renewcommand{\arraystretch}{1.0}
\end{center}
\caption{Dictionary between different notations
for the 6 selected ``highlights'' in the angular distribution of $B\to K^*\mu^+\mu^-$
and Wilson coefficients they are most sensitive to.}
\label{tab:obs}
\end{table}

Since different notations and conventions exist for the numerous $B\to K^*\ell^+\ell^-$ observables, table~\ref{tab:obs} provides a dictionary between the notation used in this report and a selection of other theory and experimental papers. It also lists the Wilson coefficients which, if modified by NP, would have the biggest impact on the observable in question.

\section{Conclusions}

\noindent
Rare $B$ decays continue to be valuable probes of physics beyond the SM. In the current early phase of the LHC era, exclusive modes with muons in the final state are among the most promising decays.

The $B_s\to\mu^+\mu^-$ decay is likely to be observed before the end of 2012 \cite{LHCbtalk}. If an enhancement beyond $10^{-8}$ is observed, this will be a clear indication of scalar operators as are present in two-Higgs-doublet models or the MSSM at large $\tan\beta$. But also moderate enhancements below a factor of 2 are predicted in many models and should be experimentally accessible. Regardless of whether $B_s\to\mu^+\mu^-$ will be discovered at its SM value or not, the companion decay $B_d\to\mu^+\mu^-$ allows a powerful test of the MFV principle by means of eq.~(\ref{eq:sd}).

$B\to K^*\mu^-\mu^-$ is more challenging experimentally as well as theoretically but offers many observables sensitive to NP. Theoretically, the challenges include calculating hadronic form factors and non-factorizable corrections and estimating uncertainties e.g. due to unknown power corrections and duality violations. Although some, possibly irreducible, sources of uncertainty remain, progress has been made recently, in particular in the high-$q^2$ region. At low $q^2$, 6 promising observables have been discussed here. four of them can be obtained from one-dimensional angular distributions; two of them are sensitive to right-handed currents; two are sensitive to CP violation.

In the coming decade, experiments will once more scrutinize the CKM description of flavour and CP violation in rare $B$ decays. Whether deviations from the SM expectations will be found or not, this effort will teach us a lot about the physics of flavour.

\section*{Acknowledgments}

\noindent
I thank Thorsten Feldmann for useful comments on the manuscript.
This work was supported by the EU ITN ``Unification in the LHC Era'', contract PITN-GA-2009-237920 (UNILHC).

\end{document}